\documentstyle[aps,preprint]{revtex}
\newcommand{\avg}[1]{\langle{#1}\rangle}
\newcommand{\req}[1]{(\ref{#1})}
\newcommand{\beq}{\begin{equation}}
\newcommand{\eeq}{\end{equation}}
\newcommand{\beqas}{\begin{eqnarray*}}
\newcommand{\eeqas}{\end{eqnarray*}}
\begin{document}
\title{Scaling Behavior in the Stable Marriage Problem}
\author{Marie-Jos\'e Om\'ero, Michael Dzierzawa\thanks{present address:
Institut f\"ur Physik, Universit\"at Augsburg, D-86135 Augsburg, Germany},
Matteo Marsili, and Yi-Cheng Zhang}
\address{Institut de Physique Th\'eorique, 
Universit\'e de Fribourg, CH-1700\\
~~\\
~~\\
Short Title: {\bf Scaling in the Stable Marriage Problem}\\
~~\\
~~\\
}
\author{PACS: 05.20.-y, 01.75.+m, 02.50.Le}
\date{\today}
\maketitle

\newpage

\begin{abstract}
We study the optimization of the stable marriage problem. 
All individuals
attempt to optimize their own satisfaction, subject to mutually 
conflicting constraints.
We find that the stable solutions are generally not the 
globally best
solution, but reasonably close to it. All the stable 
solutions form a special
sub-set of the meta-stable states, obeying interesting 
scaling laws.
Both numerical and analytical tools are used to derive
our results. 
\end{abstract}
\newpage

\section{Introduction}

Optimization problems have become an interesting
area of research in statistical physics. They usually 
require to find 
the minimum (or minima) of a global quantity (e.g.
Hamiltonian). Spin-glasses are just one 
of the current well-known examples. In the context 
of social science or economy, decision-makers are 
individuals or companies. They have their own rather 
selfish goals to optimize, which are often conflicting.
This situation, which is typical in game theory, 
cannot be described by a global Hamiltonian 
\cite{Fud91,MZ97}. One of its simplest realizations
is in the classical stable marriage problem 
\cite{GS62,Knu76,GusIrv89}. Despite its simple definition 
the solutions have a very rich structure.

The stable marriage optimization problem does not 
require the "energy" to be the smallest possible, 
but that the resulting state be stable against
the egoistic attempts of individuals to lower their
own ``energy''. This new concept of equilibrium is
typical in game theory where one deals with a number
$N$ of distinct agents, each of whom is trying to 
maximize his utility at the same time. It is not true in general
that the state with the largest total utility will be 
an equilibrium state. Indeed it is possible that in such
a state an agent would benefit from making an action 
which increases his utility at the expense of others. 
This leads to the concept of Nash equilibrium 
\cite{Nas50}, which is a state characterized by the 
stability with respect to the action of
any agent. In other words, a state is stable if any 
change in an agent's strategy is unfavorable for himself.

The marriage problem  
describes a system where two sets of $N$ persons
have to be matched pairwise. We shall assume that these
sets are composed of men and women who are to be 
married. Clearly the marriage problem is applicable in 
many different
contexts where two distinct sets have to be matched 
with
the best satisfaction. For instance, one can consider
$N$ applicants facing $N'$ jobs.
Each applicant has a preference list of jobs and each
job-owner ranks the applicants in order of preference.
For the convenience of presentation we exclusively use 
the paradigm of marriage between men and women. 
Suppose the men and the women in the two sets know 
each other well.
Based on his knowledge each man establishes a wish-list of 
his desired women,
in the descending priority order, i.e. on the top of his 
list is his dream girl;
the bottom is a mate whom he has to marry only in the 
worst case when all the other
women reject him. The women do exactly the same to 
the men.
Note that in the convention of this model, everybody must 
marry.

In our model, we make the further simplifying assumption that 
each person's satisfaction depends on the rank
of the partner he/she gets to marry. Thus the rank can be seen
as a cost function. If the top choice is attained, the cost is 
the least, the bottom choice has the highest cost.
Two men may happen to put the same woman  as their 
top choice, or two women may happen
to prefer the same man. There are necessarily 
conflicting wishes which
give a special complexity to the problem.
We shall deal here with the case in which 
each person's wish-list 
is randomly and independently established. 

\section{The Model}

In order to set up the notation
let us look at the following example of three men 
and three women. The preference lists for all 
persons are shown below.

\vspace{0.5cm}

\noindent
\hspace*{1cm} 1: 1 2 3	\hspace{3cm}		   1:  3 1 2 \\
\hspace*{1cm} 2: 1 3 2	\hspace{3cm}	           2:  1 2 3 \\
\hspace*{1cm} 3: 2 3 1	\hspace{3cm}	           3:  3 2 1 \\

Men's preference lists ~~~~~ Women's preference lists
\vspace{0.5cm}

The detailed cost for each person can be set observing 
that, e.g. if man 1 marries woman 2 his cost is
2 since she is in the second position of his preference
list and, vice versa, the cost for woman 2 is 1. If man 2 
marries woman 1, his cost is 1 and if man 3 marries woman
3 his cost is 2. The costs of women 1 and 3
in these cases are 3 and 1, respectively.

It is convenient
to introduce a representation of the lists in terms
of rankings: We define the matrices $F$  and $H$
for women and men respectively, such that $f(w,m)$
denotes the position of a man $m$ in the list of  a woman $w$.
Equivalently $h(m,w)$ yields the rank of a woman $w$ in
$m$'s list. Rank plays here a similar role to
energy in statistical mechanics, so we shall
frequently refer to $f(w,m)$ or $h(m,w)$ as energies.
A realization of the preference lists 
is also called an instance.
A matching is a set of $N$ pairs
${\cal M}=\{(m_i,w_i),\,i=1,\ldots,N\}$, and
there are $N !$ possible matchings in an instance of size $N$.

The problem is to find a {\em stable} matching
${\cal M} =\{(m_i,w_i)\}$
such that one cannot find a man $m_i$ and a woman $w_j$ who are
not married ($i\neq j$) but would {\em both}  prefer
to marry each other rather than staying with their respective
partners $w_i$ and $m_j$. Such a couple is called a
{\em blocking pair}.
A blocking pair $(m_i,w_j)$ is then such that
$f(w_j,m_i)<f(w_j,m_j)$ {\em and} 
$h(m_i,w_j)<h(m_i,w_i)$.
If no such pair exists, the matching
is called stable.

One can calculate the energy per person, for 
women and men in a given  matching as
\[
\epsilon_F({\cal M})=\frac{1}{N}\sum_{i=1}^N f(w_i,m_i),~~~~~
\epsilon_H({\cal M})=\frac{1}{N}\sum_{i=1}^N h(m_i,w_i)
\]
and the energy $\epsilon({\cal M})=\epsilon_F({\cal M})+
\epsilon_H({\cal M})$ per couple.
Here and in the following the subscripts $F$ and $H$ stand for 
women and men, respectively.

\section{The Gale-Shapley algorithm}

A lot of work has been spent in developing fast 
computer algorithms to find all stable matchings for a given 
instance of size $N$ \cite{Knu76,GusIrv89}. These algorithms are based 
on the classical Gale-Shapley (GS) algorithm  \cite{GS62} which
assigns the role of proposers to the elements of 
one set,
the men say, and of judgers to the elements of the 
other. 

The {\em man-oriented} GS algorithm starts 
from a man $m$ making a proposal to the first
woman $w$ on his list. If she accepts they get married, 
if she
refuses $m$ goes on proposing to the next woman on his 
list. 
$w$ accepts a proposal when either she is not engaged or
she is engaged with a man $m'$ worse than the one 
proposing ($m$). 
In the latter case, $m'$ will have to go on proposing to 
the
woman following $w$ on his list. When all men have run 
through
their lists proposing until all women are married, the 
algorithm
stops and the matching thus reached is a stable 
matching.

As an illustration let us consider the example of 
the preceding paragraph.
The man-oriented GS algorithm goes as follows:
man $1$ proposes to woman $1$ who accepts and they form the
pair $(1,1)$. Then $2$ proposes to $1$, but she 
refuses.
So man $2$ proposes to woman $3$ and they get married. 
Finally man $3$ happily marries woman $2$. This results
in the matching ${\cal M}_{H}=\{(1,1),(2,3),(3,2)\}$. 

The GS algorithm can be run reversing the roles 
(woman-oriented) to yield the  woman-optimal
stable matching. In our example, this leads to 
${\cal M}_{F}=\{(1,1),(2,2),(3,3)\}$. 

The energies for men and women in these matchings are
$(\epsilon_H,\epsilon_F)=(4/3,7/3)$ and $(6/3,5/3)$ 
respectively for ${\cal M}_H$ and ${\cal M}_F$. 
As seen in this example who proposes
is always better off than who judges.

It can be shown \cite{Knu76} that the man-oriented 
GS algorithm yields the 
man-optimal stable matching in the sense that no man 
can have a better partner in any other stable matching. 
It is rather surprising that though
men get rejections nearly all the time 
and just one positive answer, they are far better off 
than women. On the other hand 
women who take the pleasure by saying no to almost all 
the suitors except one who is best among her suitors, 
will end up in a marriage
that is the worst among all possible stable ones. The 
lesson is that the person
who takes initiative is rewarded.

In order to quantify more precisely this statement,
it is enough to observe that {\em i)} once a woman
is first engaged, she will remain engaged (eventually 
with different men) forever, and that {\em ii)} the
total energy for men in the man-optimal GS equals 
the total number of proposals men need to make to
marry all women. Proposals by men define an 
intrinsic time in the algorithm. Imagine that at 
time $t_k$ (i.e. after $t_k$ proposals) the
$k^{\rm th}$ woman gets engaged. 
In view of the randomness of the preference lists,
the probability 
that the next proposal is addressed to one of the $N - k$
free women is $1 - k / N$. On average, men will 
need $N/k=\avg{t_{k+1}-t_k}$ proposals to engage 
one more woman. Since the total energy for men
is $N\epsilon_H=t_N$, we find
\beq
\epsilon_H({\cal M}_{H})=\sum_{k=1}^N\frac{1}{k} \approx \log N + C 
\label{ehGS}
\eeq
where $C = 0.5772\ldots$ is Euler's constant. 
Taking into account that men do not propose twice to the same woman
yields a correction of ${\cal O}((\log N)^2 / N)$ to eq. (1).
On the average each woman receives $\log N + C$ proposals of men 
who are randomly
distributed between $1$ and $N$ on her preference list.
Keeping only the best proposal the women arrive at an energy
of the order of $\epsilon_F({\cal M}_{H})\simeq N/\log N$
which is much larger than the corresponding result (1) for the men.

\section{Stable Matchings in the large $N$ limit}

Coming back to our $N=3$ example, it is also interesting to
note that none of the two stable states we found is
the one with the smallest energy, the ``ground state''.
Indeed the state ${\cal M}_0=\{(1,2),(2,1),(3,3)\}$
has $(\epsilon_H,\epsilon_F)=(5/3,5/3)$, and a total energy 
$\epsilon({\cal M}_0)=10/3$ which is lower than those of the 
other two states. This state is however unstable. 
Indeed there is one blocking pair $(1,1)$.

This simple example already shows some interesting 
features of the stable marriage problem: 
{\em 1)} minimum energy i.e. maximum global satisfaction
does not imply stability (and vice versa) and {\em 2)}
there can be more than one stable matching (the GS 
algorithm guarantees that there is always at least one stable 
matching).

In order to decrease the total energy of a given stable
matching, it would be necessary that some individuals pay 
the price of accepting a worse partner than the one with whom they 
are actually married. But in absence of a supervising body (like
government or parents) they do not consider to help others by switching.
The inherent selfishness and conflicting optimizations
in the stable marriage 
problem lead to stable solutions that are not globally optimal. 
%{\em When 
%there are more members this feature remains; their optimization attempts
%compromise to achieve self-organised stable solutions.} (?????)
 
It is known that the average number of stable states
in an instance of size $N$ is proportional to $N\log N$ \cite{GusIrv89}.
Starting from the man-optimal Gale-Shapley solution all 
other stable matchings
can be obtained by performing cyclic exchange processes 
\beq
(\mu_1,\omega_1),\ldots,(\mu_r,\omega_r) 
\rightarrow (\mu_1,\omega_2),\ldots,(\mu_r,\omega_1)
\label{rotation}
\eeq
within a properly chosen group of pairs of
men $\mu_i$ and women $\omega_i$.
These exchange processes are called rotations. 
The number $r$ of pairs involved in a rotation 
can be regarded as the "distance" between the two stable 
matchings 
connected by this rotation. 
Efficient algorithms exist for finding all rotations
and therefore all stable matchings in a given instance \cite{GusIrv89}.
In this way it has been found that stable states are
organized in very peculiar graph-theoretical structures
\cite{GusIrv89}. 

We shall first analyze the statistics of stable states and
then return to the concept of rotations.

It is possible to derive a general relation between
the energies of men and women in the stable states.
Our aim is to evaluate the energy of men, given that
of women in a stable state. We can then assume that
a stable matching, in which woman $w$ have energy 
$\epsilon_w$, exists. In order to find the stable
matching we consider the ``hypothetical'' situation in 
which,  for some reason, the women know that they 
can reach a stable matching where woman $w$ has 
energy $\epsilon_w$. Knowing this, the best
strategy of women, becomes that of refusing
all propositions from men ranking higher than 
$\epsilon_w$. The best strategy for men, on the
other hand, remains that of the GS algorithm.
The dynamics of this modified GS algorithm will 
clearly reach the stable state foreseen by women:
Each man $m$ makes his proposals sequentially until 
he hits a woman $w$ such that $f(w,m)\le\epsilon_w$.
In order to compute the average man energy,
consider a man $m$ and, in order to simplify the
notations, let his wish list be $h(m,w)=w$, for
$w=1,\ldots,N$\footnote{This can always be achieved 
with a permutation of women indices. 
Clearly $h(m',w)\neq w$ for $m'\neq m$, in general.}
His proposal to the $w^{\rm th}$ woman, will
fall randomly within $1$ and $N$ in the rankings
of woman $w$ and, according to the above strategy,
it will be accepted with a probability 
$\epsilon_w/N$. Otherwise, if it is refused, man
$m$ will consider the $w+1^{\rm st}$ woman. 
The probability that this man will get his 
$q^{\rm th}$ choice is then
\beq
P_H(q) = \prod_{w=1}^{q-1}
\left(1-\frac{\epsilon_w}{N}\right)
\frac{\epsilon_q}{N}.
\label{p1}
\eeq

We can now take the average on realizations of the above
equation. Under the assumption that $\epsilon_w$ are independent 
random variables with average $\epsilon_F$, we find
\beq
P_H(q) = 
\left(1-\frac{\epsilon_F}{N}\right)^{q-1}
\frac{\epsilon_F}{N}.
\label{p2}
\eeq
From this we can compute the average energy for the
$m^{\rm th}$ man $\epsilon_H=\sum q P_H(q)$. This leads to
the relation
\beq
\epsilon_H\,\epsilon_F=N.
\label{ehef}
\eeq
In other words, {\em in stable matchings, for random 
instances of the marriage problem, the energies for
men and women are inversely proportional.}

Of course, reversing the roles of men and women, 
one finds the same conclusion eq. \req{ehef} and 
a distribution $P_F(q)\simeq \exp(-q/\epsilon_H)/
\epsilon_H$ of women energies which depends on 
$\epsilon_H$. 

It is now possible, returning to the assumption of
independence of the $\epsilon_w$'s, to show that
only a weak correlation exists so that equations
\req{p2} and \req{ehef} are exact in the limit $N\to\infty$.
In order to do this we can run the above 
argument in its women-oriented version (with women
proposing) to derive the 
joint probability distribution of the energies of two 
women. With independent men energies, it is clear 
that unless two women propose to the same man, 
there will be no correlation between their energies.
The probability that both propose to the same
man is of the order of $\epsilon_F^2/N^2$, 
which implies a weak correlation of the form 
\beq
\avg{\delta \epsilon_j\delta \epsilon_i}=
\frac{c}{N}\avg{\epsilon_i}^2
\label{corr}
\eeq
among energies.
This clearly holds both for women and for men
{\em under the assumption} of the independence
of men or women energies, respectively.
It can be easily seen that women (men) energies
are also weakly correlated, as in eq. \req{corr},
if men (women) energies are weakly correlated.
We therefore conclude self-consistently that energies 
are weakly correlated in stable matchings.

In presence of a weak correlation of the form
\req{corr}, the same procedure, from
eq. \req{p1} to eq. \req{ehef} leads to 
$\epsilon_H\epsilon_F=N(1+4c/N)$.
Therefore eq. (\ref{ehef}) is exact in the limit $N\to\infty$.
We found numerically that the constant $c$ is generally
negative ($c\simeq -0.3$ in man optimal states). This 
correlation is similar to the one occurring 
among $N$ variables whose sum is constrained.

Our numerical results indicate that the relation \req{ehef}
is already satisfied approximatively for a rather small 
number of pairs. Fig. 1 shows $\log \epsilon_H$
as a function of $\log \epsilon_F$ for all stable 
matchings found
in systems of size $N = 50,100,200,500,1000$. The 
asymptotic result
$\log \epsilon_H + \log \epsilon_F = \log N$ is 
indicated by lines.
The points on the left and on the right of each set of 
data correspond
to the man and the woman optimal GS solutions, 
respectively.

Fig. 2 shows that the distribution of individual energies 
in a stable matching agrees very well with the predicted
exponential behavior eq. \req{p2}.

Although, as shown by the GS algorithm, stable matchings 
can be very asymmetric regarding the energy
of men and women, there is nevertheless a minimum energy
of order ${\cal O}(\log N)$ that cannot be reduced 
further without losing stability.
Stable solutions where either men or women possess an 
average energy of ${\cal O}(1)$ are not possible.  

\section{Dynamics between stable states}

Rotations play a central role in the algorithm
which finds all stable states. As for the GS, this
algorithm has a man--oriented version and its
woman--oriented counterpart. As mentioned a 
rotation is a cyclic permutation of partners 
within a subset of persons in a stable matching ${\cal M}$, 
which allows to reach a new stable matching ${\cal M}'$.
In the man--oriented algorithm, to which we shall
restrict attention, the execution of a rotation 
raises the energy of any man involved, and 
lowers the energy of the corresponding woman. 
In this way, starting from the man--optimal state,
the execution of rotations, in all possible orders
\footnote{A rotation, like the one in eq. (2), can be
executed on a stable state only if it is exposed,
i.e. only if each man $\mu_i$ is paired, in that state, 
with the woman $\omega_i$ specified in eq. (2), for 
$i=1,\ldots, r$.},
allows to reach sequentially all other stable states
until the woman optimal one. This process, therefore,
runs through the set of stable states shown in Fig. 1
from top (the man--optimal state) to bottom (the 
woman--optimal state). 

We can understand this process with a generalization of 
the GS dynamics. Imagine that in a stable state ${\cal M}=
\{(m_i,w_i)\}$, a woman $\omega_1$, for some reason 
divorces. This event leaves an unpaired man $\mu_1$,
and the GS dynamics starts again: $\mu_1$ will run 
through his list making proposals to the women following 
$\omega_1$ until he finds a new woman $\omega_2$
which prefers him to her partner $\mu_2$. This 
will put man $\mu_2$ in the same situation as 
man $\mu_1$ before. The process will continue, 
involving other men $\mu_i$ and women $\omega_i$, until 
a man $\mu_r$ will make a proposal to woman $\omega_1$.
Under the GS dynamics, this proposal will be accepted,
because $\omega_1$ is free. It might happen that 
the new partner of $\omega_1$ is better than the
one she left $f(\omega_1,\mu_r)<f(\omega_1,\mu_1)$.
In this case the state thus reached will again be
stable. The dynamical process described above, 
exactly represents in this case, the execution of the 
rotation \req{rotation}.

On the other hand, if, for woman $\omega_1$, $\mu_r$ is 
a worse partner than $\mu_1$, the state will be unstable.
Indeed $(\mu_1,\omega_1)$ constitute in this case a 
blocking pair: both of them would indeed prefer to
get together again than to be married with $\omega_2$
and $\mu_r$ respectively. We can, in this case,
regard the above process as a ``virtual'' process
that does not lead to a new stable state and
that leaves the state unchanged. 

Note that the probability that the next proposal $\omega_1$ 
receives is better than the one she holds in the 
stable state is $f(\omega_1,\mu_1)/N\simeq\epsilon_F/N$.
$\epsilon_F \sim \sqrt{N}$ implies that the probability
that any woman improves her situation with a divorce
is very small $O(1/\sqrt{N})$. With such a
small probability divorce is very risky for any woman
under the strategies fixed by the GS algorithm
(Note, on the other hand, that there will be $O(\sqrt{N})$
women $\omega_1$ for which the above process leads 
to a new stable state). 

This dynamics motivates a deeper study of the statistics
of rotation lengths. Indeed we can understand the 
execution of a rotation as the response of the system
to the small perturbation which causes the first divorce.
While the response is generally linear in equilibrium
statistical systems, we shall see that a small
perturbation on a stable marriage can cause a large
response, i.e. a large change in the system. This is
reminiscent of the behavior of self organized
critical systems. The statistics of rotation lengths
is also interesting to understand the ``geometrical''
nature of the organization of stable states. Indeed
the rotation length $r$ measures the ``distance''
between two stable states, i.e. the number of marriages
which differ in the two matchings.

We found that the normalized distribution of the length 
$r$ of a rotation satisfies
\beq
P(r,N)=
\frac{1}{r_0(N)}
\rho\left[
\frac{r}{r_0(N)}
\right]
\eeq
with the typical rotation size scaling with $N$ as
$r_0(N)\sim \sqrt{N}$.

The scaled distributions are shown in Fig. 3. All data 
collapse on a single curve which is remarkably well fit by a
Gaussian $\rho(x) \simeq \frac{\sqrt{\pi}}{2}\exp (-x^2)$ (line).
This scaling behavior can be understood considering
the above mentioned extension of the GS algorithm.
Note indeed that the number of men involved in 
the process is $r$. Using arguments similar to the ones leading
to eq. \req{ehef}, one sees that each man 
typically needs an additional $N/\epsilon_F$ 
proposals. So the total number of proposals
received by $\omega_1$ is $r/\epsilon_F$. 
The length $r$ of the rotation is obtained
imposing that $\omega_1$ receives $\sim 1$ proposition.
This implies that $r$ will be typically of the 
order of $\epsilon_F\sim\sqrt{N}$. It also implies
that the men's energy difference between the two
states is $d\epsilon_H\simeq r/\epsilon_F$, which
is of order one. This agrees, apart from logarithmic
corrections, with the observation 
\cite{GusIrv89} that there are $\sim N\log N$ stable
matchings.

\section{Globally optimal solution}

While there exist powerful numerical algorithms to 
obtain all stable
matchings in a systematic way it is a much harder 
problem to find
the ground state, i.e. the matching with minimal total 
energy.
For an analytical approach it is convenient to introduce 
the random variable
\beq
x(m,w) = \frac{1}{N}\left[h(m,w) + f(w,m) \right]
\eeq
which is the normalized energy associated with the 
formation of the
pair $(m,w)$. Since we are no longer interested in 
stability considerations 
it is not necessary to distinguish between the 
energy of men and women. In the large $N$ limit 
$x(m,w)$ can be
treated as a continuous random variable with 
distribution
$\rho(x) = \min(x,2-x)$ for $0 < x < 2$. The problem of 
finding
the minimum of
\beq
\epsilon({\cal M}) = \sum_{i=1}^N x(m_i,w_i)
\eeq
over all matchings ${\cal M}$ reduces to the bipartite matching
problem which has been solved by M\'ezard and Parisi \cite{Mez85}
using the replica technique for the disorder average. 
Following their approach we obtain
\beq
\epsilon_{\min} = 1.617 \sqrt{N}.
\eeq
On the other hand, minimization of $\epsilon({\cal M})=
\epsilon_F({\cal M})+\epsilon_H({\cal M})$ subject to the
stability condition, i.e. to eq. \req{ehef}, yields
\beq
\epsilon_{\min}^{stable} = 2 \sqrt{N}.
\eeq
Thus giving up the constraint of stability allows for a 
reduction of energy
by 19 percent.

%From the scaling behavior in Fig.2 we see that the 
%extreme cases
%for both men and women oriented GS algorithms the total 
%cost to
%the society is maximum, thus undesirable. This is 
%because that the total cost
%is the sum of two elements whose product is constant. 
%On the other
%hand around the mid-point on Fig.2 the stable solutions 
%are also society
%friendlier, the total cost being the minimum. 

%This can be achieved and the two
%general scenarios: 1) every person has $50\%$ chance of 
%being positive 
%regardless his/her sex, in this way about half of the 
%men and women take 
%initiatives. 2) Each 
%single person cannot be labelled 
%positive or passive,
%rather he/she can randomly ($50\%$ chance again) switch 
%from proposal making and 
%proposal taking, e.g. if he/she is the passive  mood, he 
%just retain the best
%suitor, but if the mood turns positive he/she will make 
%his/her own proposal,
%running down the wish-list, while still keep the best 
%candidate as fiancee/fiance.
%This continues until everybody is married.

%Unsymmetrical cases can be also envisaged. In Fig. 2 we 
%plot also
%the cost distribution when say 2/3 of the men are 
%positive and so are 1/3 of
%the women. Starting with different initial conditions we 
%see the solutions
%(by the points) are scattered...

\section{Conclusions}

We have investigated the classical stable marriage problem
which despite its simple definition contains the full complexity
of highly frustrated systems like spin glasses. In contrast to the 
traditional examples of statistical mechanics where the 
dynamics of the system is governed by a single global quantity, 
the Hamiltonian, the stable marriage
problem fits more naturally into the framework of game theory where
the concept of Nash equilibrium plays the central role. 
The game-theoretical definition of stability  
leads to the somewhat paradox result that although all individuals
do whatever they can in order to maximize their personal benefit
the resulting stable states are not the globally best solution.
In the large $N$ limit, we found stable states with total energies 
ranging from
$\epsilon_{max}^{stable} = N / \log N$ to 
$\epsilon_{min}^{stable} = 2 \sqrt N$,
whereas the globally best solution has $\epsilon_{min} = 1.617 \sqrt N$.
Within a stable matching the distribution of individual energies,
say for the men, 
is decaying exponentially where the decay constant is determined
by the mean energy of the women, and vice versa.
As a consequence, the mean energies of women and men satisfy the simple
relation $\epsilon_H\,\epsilon_F = N$.
We also studied the distribution of distances between stable matchings
in an instance of size $N$. This distribution turned out to be a universal
function when the distances are scaled with the typical rotation length
$r_0 \sim \sqrt N$. Introducing a simple dynamics, this result also
implies that the system is characterized by a 
non--linear response to perturbation similar to the one 
observed in self organized critical systems.

There are still many open questions to be investigated
in this problem. It would be interesting to compare the 
distribution of individual energies in the ground state,
i.e. the globally best solution, with the one we found 
in stable matchings. The latter, as shown, are well
described by independent variables with a common
distribution. On the other hand one expects that, 
in the ground state, they are much more
strongly (anti-) correlated and that individual energies 
can fluctuate much more wildly (see e.g. \cite{M96}).

Another interesting question is how many blocking 
pairs there are in the ground state since this would be a
measure for its degree of instability. A very rough 
argument suggests that this number is of order 
$\epsilon_{\min}^2/4\sim N$.

%To this end a 
%powerful numerical optimization
%method that allows to find the ground state in reasonably large systems
%has to be developed. 
We are also presently investigating a generalization of the
model where the assumption 
that the preference lists 
of different individuals are uncorrelated is 
replaced by a more realistic hypothesis.

\section{Acknowledgements}

This work was supported in part by the Swiss National 
Foundation under grant 20-40672.94/1.

\begin{figure}
\caption{Average energy of men $(\epsilon_H)$ and women $(\epsilon_F)$
for all stable matchings in systems of size $N = 50,100,200,500,1000$
(from left to right)
on a double logarithmic scale. The analytic result of 
Eq. (5) is indicated
by lines. The $\Diamond$ point below each line 
corresponds to the Ground State energy.}
\label{Fig.1}
\end{figure}

\begin{figure}
\caption{Distribution of individual energies of men (open circles) and
women (full circles) for all stable matchings in an instance of size 
$N = 1000$ on a logarithmic scale. The energies of men (women)
are scaled by the their average values: $x=h(m_i,w_i)/\epsilon_H$
for men, and $x=f(w_i,m_i)/\epsilon_F$.
The solid line is the analytic result of Eq. (4).}
\label{Fig.2}
\end{figure}

\begin{figure}
\caption{Scaled distribution of rotation lengths for systems of size
$N = 100, 200, 500, 1000$. The curve is 
$\propto\exp(-x^2/2)$. }
\label{Fig.3}
\end{figure}

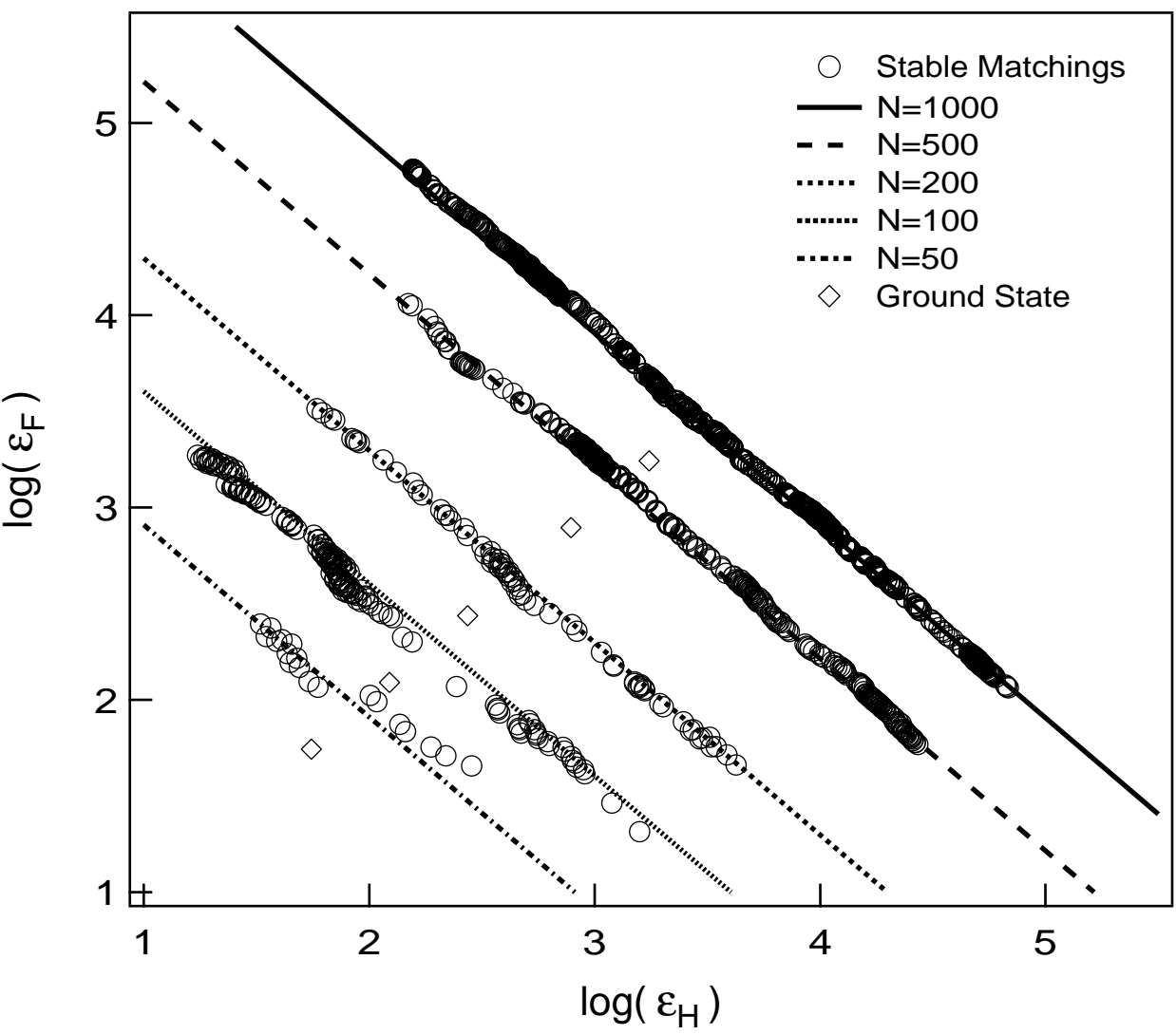
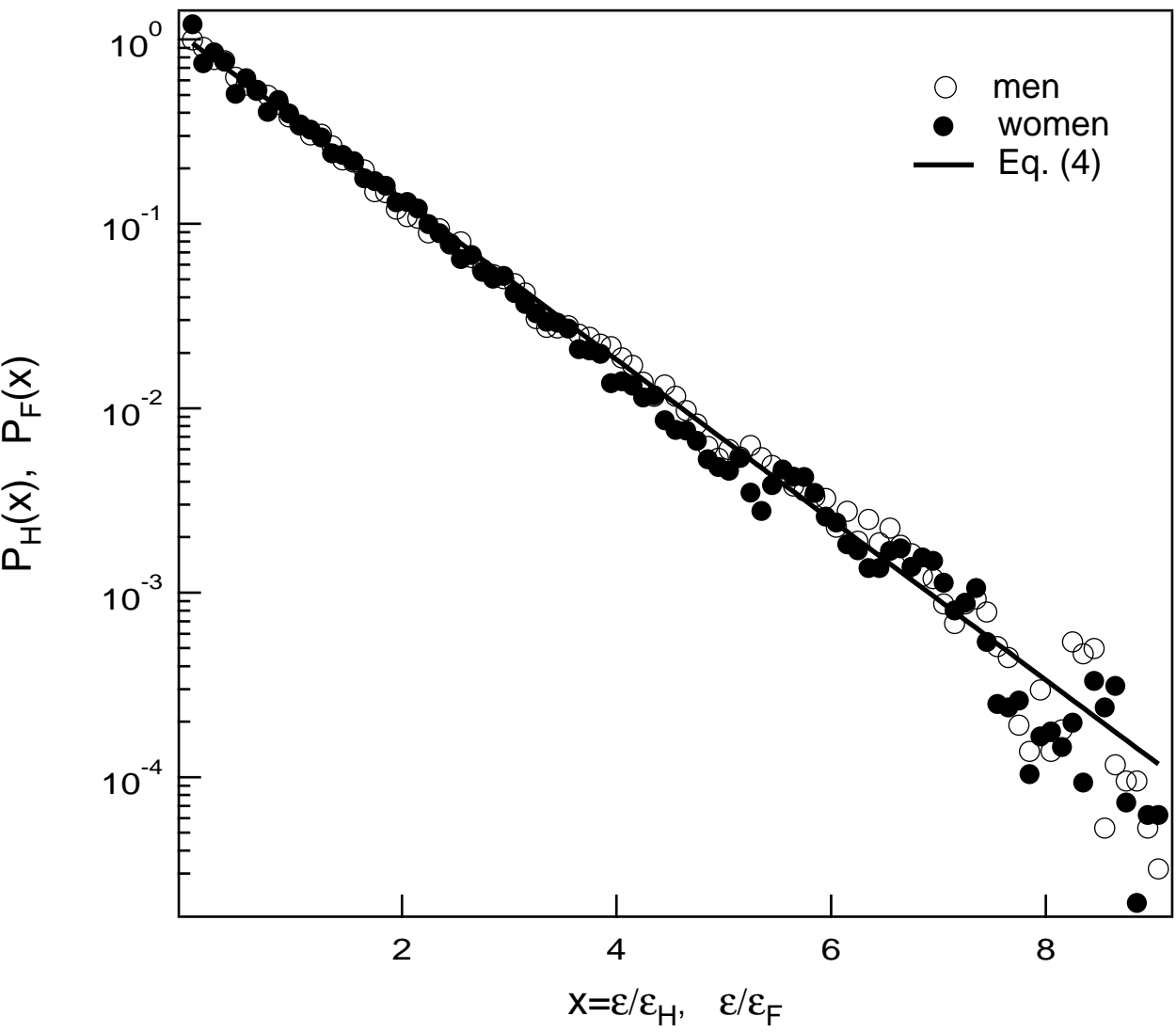
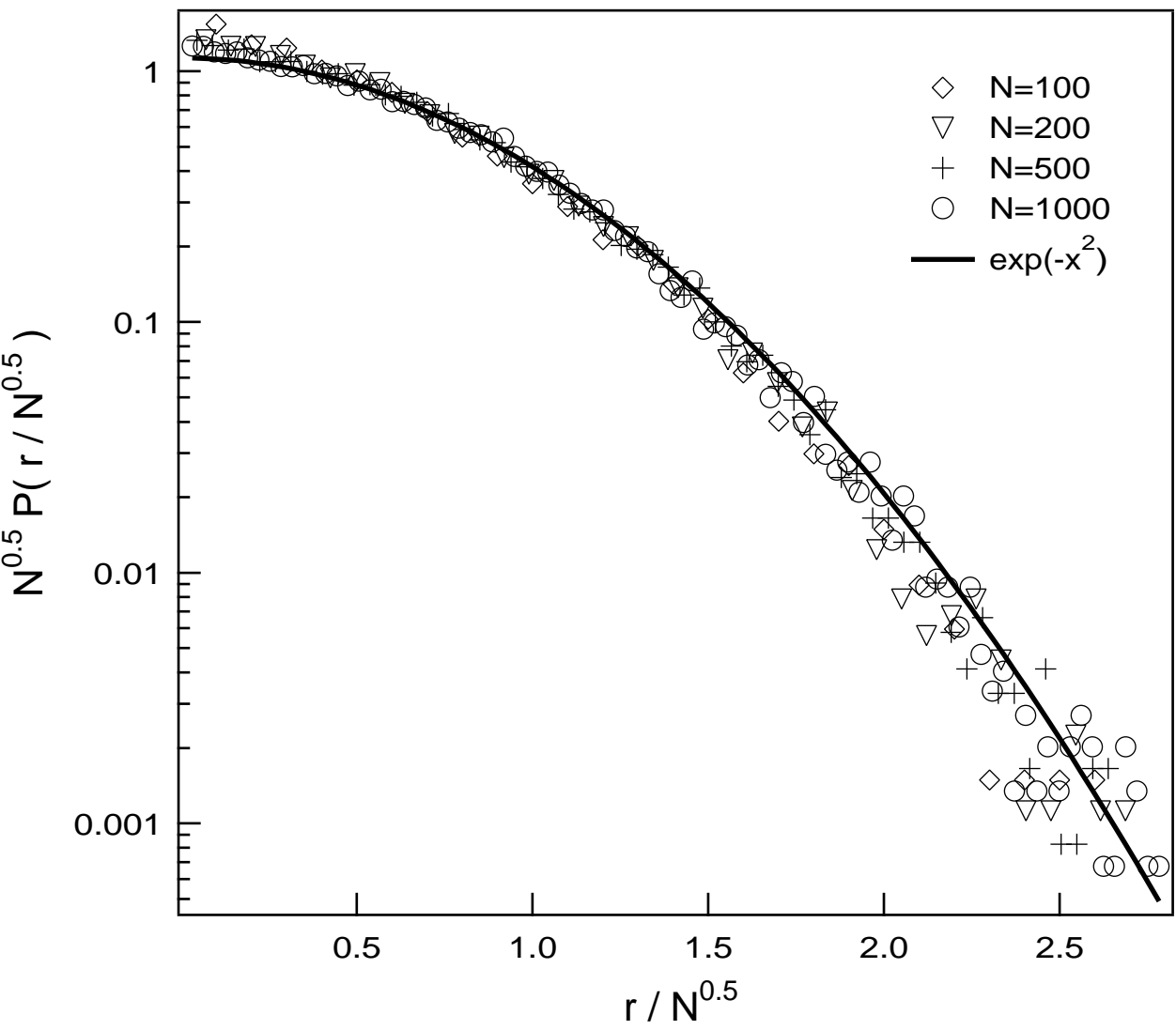
\end{document}